\documentclass[aps,reprint,superscriptaddress,longbibliography]{revtex4-1}

\usepackage{floatrow}
\usepackage{graphicx}
\usepackage{here}
\usepackage{amsmath,amssymb,amsfonts}
\usepackage{physics}
\usepackage{color}
\usepackage[T1]{fontenc}
\usepackage{textcomp}
\usepackage{hyperref}
\hypersetup{
     colorlinks   = true,
     linkcolor    = blue,
     citecolor    = blue,
     urlcolor     = blue
}

\usepackage{physics}

\DeclareUnicodeCharacter{2061}{}

\begin{document}

\title{$\mathbb{Z}_2$ Non-Hermitian skin effect in equilibrium heavy-fermions }

\author{Shin Kaneshiro}
\email{kaneshiro.shin.88a@st.kyoto-u.ac.jp}
\author{Tsuneya Yoshida}
\author{Robert Peters}
\affiliation{Department of Physics, Kyoto University, Kyoto 606-8502, Japan}
\date{\today}

\begin{abstract}
We demonstrate that a correlated equilibrium $f$-electron system with time-reversal symmetry can exhibit a $\mathbb{Z}_2$ non-Hermitian skin effect of quasi-particles.
In particular, we analyze a two-dimensional periodic Anderson model with spin-orbit coupling by combining the dynamical mean-field theory (DMFT) and the numerical renormalization group.
We prove the existence of the $\mathbb{Z}_2$ skin effect by explicitly calculating the topological invariant and show that spin-orbit interaction is essential to this effect.
Our DMFT analysis demonstrates that the $\mathbb{Z}_2$ skin effect of quasi-particles is reflected on the pseudo-spectrum. 
Furthermore, we analyze temperature effects on this skin effect using the generalized Brillouin zone technique, which clarifies that the skin modes are strongly localized above the Kondo temperature.
\end{abstract}

\maketitle

\section{introduction}
Non-Hermitian systems attract much attention because they exhibit novel phenomena
\cite{ashida2020nonHermitianPhysics,bergholtz2021ExceptionalTopologyNonHermitian}.
The interaction between the system and the environment breaks the Hermiticity of the Hamiltonian and enriches the band structure through complex energy.
The complex band structures provide novel aspects to the topology that do not have Hermitian counterparts, such as the exceptional point (EP)
\cite{ruter2010ObservationParityTime, regensburger2012ParityTimeSynthetic, zhen2015SpawningRingsExceptional, zhou2018ObservationBulkFermi,miri2019ExceptionalPointsOptics,zhou2019ExceptionalSurfacesPTsymmetric,
san-jose2016MajoranaBoundStates,martinezalvarez2018NonHermitianRobustEdge,
shen2018TopologicalBandTheory,rui2019PTSymmetricNonHermitian,yoshida2022FateExceptionalPoints}
and the Non-Hermitian skin effect (NHSE)
\cite{yao2018EdgeStatesTopological,kunst2018BiorthogonalBulkBoundaryCorrespondencea,
edvardsson2019NonHermitianExtensionsHigherordera,
ezawa2019NonHermitianBoundaryInterface,kawabata2019SymmetryTopologyNonHermitian,Lee2019AnatomyOfSkinModesAndTopology,luo2019HigherOrderTopologicalCorner,
borgnia2020Non-HermitianBoundaryModes,
ghatak2020ObservationNonHermitianTopology,helbig2020GeneralizedBulkBoundary,hofmann2020ReciprocalSkinEffect,kawabata2020HigherorderNonHermitianSkin,kawabata2020NonBlochBandTheory,
li2020CriticalNonHermitianSkin,longhi2020UnravelingNonHermitianSkin,okuma2020TopologicalOriginNonHermitian,okuma2020HermitianZeroModes,weidemann2020TopologicalFunnelingLight,
xiao2020NonHermitianBulkBoundary,yang2020NonHermitianBulkboundaryCorrespondence,yoshida2020MirrorSkinEffect,zhang2020CorrespondenceWindingNumbers,
kawabata2021TopologicalFieldTheory,li2021QuantizedClassicalResponse,liu2021NonHermitianSkinEffect,okuma2021NonHermitianSkinEffects,palacios2021GuidedAccumulationActive,
yoshida2021RealspaceDynamicalMean,zhang2021AcousticNonHermitianSkin,zou2021ObservationHybridHigherorder,wu2022ConnectionsOpenboundarySpectrum,zhang2022UniversalNonHermitianSkin,
okuma2023NonHermitianTopologicalPhenomena,lin2023TopologicalNonHermitianSkin}.
The former represents a band touching with a degeneracy of the eigenstates, which makes the Hamiltonian non-diagonalizable.
The latter leads to a large number of localized boundary modes (i.e., skin modes) corresponding to the non-trivial point-gap topology in the bulk.
Non-Hermiticity appears not only in open quantum systems
\cite{diehl2011TopologyDissipationAtomic,bardyn2013TopologyDissipation,rivas2013DensitymatrixChernInsulators,budich2015DissipativePreparationChern,gong2017ZenoHallEffect, yoshida2019NonHermitianFractionalQuantum,lieu2020TenfoldWayQuadratic,yoshida2020FateFractionalQuantum,haga2021LiouvillianSkinEffect,yang2021ExceptionalSpinLiquids},
but also in various classical systems, exemplified by photonic crystals
\cite{ozawa2019TopologicalPhotonics,weidemann2020TopologicalFunnelingLight,xiao2020NonHermitianBulkBoundary},
and electric circuits
\cite{imhof2018TopolectricalcircuitRealizationTopological,lee2018TopolectricalCircuits,hofmann2019ChiralVoltagePropagation,lee2020ImagingNodalKnots,zhang2022ObservationNonHermitianAggregation}.

On the other hand, strongly correlated systems (SCS) have been studied for a long time because of their exciting phenomena, such as the Mott transition, high-temperature superconductivity, and the Kondo effect. 
In particular, in $f$-electron systems, where strongly localized electrons scatter itinerant conduction electrons, the Kondo effect occurs and causes dramatic changes in the physical quantities at low temperatures\cite{hewson1993KondoProblemHeavy}.

Interestingly, recent studies have highlighted the non-Hermitian aspects of SCS
\cite{okugawa2019TopologicalExceptionalSurfaces,michishita2020EquivalenceEffectiveNonHermitian,xu2017WeylExceptionalRings,yoshida2018NonHermitianPerspectiveBand,
budich2019SymmetryprotectedNodalPhases,kawabata2019ClassificationExceptionalPoints,kimura2019ChiralsymmetryProtectedExceptional,
okuma2019TopologicalPhaseTransition,yoshida2019SymmetryprotectedExceptionalRings,michishita2020RelationshipExceptionalPoints,nagai2020DMFTRevealsNonHermitian,
wojcik2020HomotopyCharacterizationNonHermitian,delplace2021SymmetryProtectedMultifoldExceptional,
mandal2021SymmetryHigherOrderExceptional,yang2021FermionDoublingTheorems}.
In SCS, physical quantities observed in experiments are often expressed by single-particle Green's functions,
which include correlation effects by the self-energy.
In general, the self-energy is a complex number, and its imaginary part corresponds to a finite lifetime of the quasi-particles due to the scattering by other particles.

Therefore, the single-particle spectrum of SCS may host EPs and their symmetry-protected variants
\cite{xu2017WeylExceptionalRings,yoshida2018NonHermitianPerspectiveBand,budich2019SymmetryprotectedNodalPhases,kawabata2019ClassificationExceptionalPoints,kimura2019ChiralsymmetryProtectedExceptional,okuma2019TopologicalPhaseTransition,yoshida2019SymmetryprotectedExceptionalRings,michishita2020RelationshipExceptionalPoints,xu2017WeylExceptionalRings,yoshida2018NonHermitianPerspectiveBand,budich2019SymmetryprotectedNodalPhases,kawabata2019ClassificationExceptionalPoints,kimura2019ChiralsymmetryProtectedExceptional,okuma2019TopologicalPhaseTransition,yoshida2019SymmetryprotectedExceptionalRings,nagai2020DMFTRevealsNonHermitian,wojcik2020HomotopyCharacterizationNonHermitian,delplace2021SymmetryProtectedMultifoldExceptional,mandal2021SymmetryHigherOrderExceptional,yang2021FermionDoublingTheorems}.
In addition, temperature effects are also discussed for heavy-fermion systems where EPs emerge around the Kondo temperature \cite{yoshida2018NonHermitianPerspectiveBand,nagai2020DMFTRevealsNonHermitian,michishita2020RelationshipExceptionalPoints}.

In contrast to EPs and their symmetry-protected variants, the NHSE has not been sufficiently explored \cite{yoshida2021RealspaceDynamicalMean}.
In particular, it remains unclear which correlated systems exhibit the NHSE protected by the symmetry of the many-body Hamiltonian. Besides,
temperature effects on such a symmetry-protected NHSE have not been discussed.

We here demonstrate that a heavy-fermion system in equilibrium exhibits the $\mathbb{Z}_2$ NHSE protected by the time-reversal symmetry of the many-body Hamiltonian.
In particular, we analyze the two-dimensional periodic Anderson model with spin-orbit coupling using dynamical mean-field theory (DMFT) combined with numerical renormalization group (NRG).
We find that the main conditions for the appearance of the NHSE are spin-orbit coupling and dissipation caused by correlations, which create a one-dimensional nontrivial point-gap topology protected by time-reversal symmetry.
In addition to the emergence of the $\mathbb{Z}_2$ NHSE, we study finite temperature effects on this NHSE and show that this point-gap and the localization strength of the boundary states are strongly affected by the Kondo effect occurring in this model. 
Because SOC exists naturally in heavy-fermion systems, this skin effect is expected to affect the single-particle properties in many $f$-electron systems.

The rest of this paper is organized as follows:
In Sec.~\ref{setion:ModelandMethods}, we explain the model and discuss the appearance of the symmetry-protected NHSE in Sec.~\ref{section:Results}.
Finally, in Sec.~\ref{section:conclusion}, we summarize and conclude this paper.
Appendix \ref{subsec:appendix_Z2NHSE} provides a complementary explanation of this $\mathbb Z_2$ NHSE.
Appendices \ref{subsec:appendix_GBZ} and \ref{subsec:appendix_SimTrans} briefly review the non-Bloch band theory and the similarity transformation method for the symplectic class.

\section{Model and Methods}
\label{setion:ModelandMethods}
We study a periodic Anderson model with spin-orbit coupling between the $c$, and $f$ electrons on a two-dimensional (2D) square lattice \cite{michishita2019ImpactRashbaSpinorbit}.
The Hamiltonian reads
\begin{align}
    H &= H_0 + H_{\text{int}}\\
    H_0 &= \sum_{\vb*{k} \sigma} \bqty{(\epsilon_{c\vb*k} + \mu_c) c^\dagger_{\vb* k \sigma} c_{\vb* k \sigma} + (\epsilon_{f\vb*k} + \mu_f) f^\dagger_{\vb* k \sigma} f_{\vb* k \sigma}} \nonumber \\
    &+ \sum_{\vb*k \sigma} \pqty{V \sigma_0 + \vb*\alpha \cdot \vb* \sigma} \pqty{f^\dagger_{\vb* k \sigma} c_{\vb* k \sigma} + \text{h.c.}} \nonumber \\
    H_{\text{int}} &= U \sum_i n_{\mathrm{f}i\uparrow} n_{\mathrm{f}i\downarrow},
\end{align}
where
\begin{align}
    \epsilon_{a \vb*{k}} &= -2 t_{a} (\cos k_x + \cos k_y) \qc a = c,f \\
    \vb*\alpha &= \alpha(\sin k_x, \sin k_y,0).
\end{align}
Here, $c^\dagger_{\vb* k \sigma}$ ($f^\dagger_{\vb* k \sigma}$) creates a $c$ electron ($f$ electron) in momentum $\vb*{k}$ and spin direction $\sigma$. 
$\epsilon_{c/f\vb*k}$ are the dispersions of the $c$ and $f$ electrons, and $\mu_{c/f}$ are the corresponding local energies.
While $V$ is a local and spin-independent hybridization, $\alpha$ corresponds to a nonlocal spin-dependent hybridization originating in the SOC.
Finally, a local density-density interaction exists between the $f$ electrons on the same atom with strength $U$.
Throughout this paper, we set $t_c=1$, $t_f=0$, $V=0.4$, $U=2$, and $\mu_f=-U/2=-1$. 

To analyze correlation effects in this model, we use the dynamical mean-field theory (DMFT)
\cite{metzner1989CorrelatedLatticeFermions,muller-hartmann1989CorrelatedFermionsLattice,georges1996DynamicalMeanfieldTheory}.
DMFT maps each lattice site onto a quantum impurity model, which must be solved self-consistently.
From this self-consistent solution, we obtain a momentum-independent self-energy.
To calculate the self-energy for the quantum impurity model, we use the numerical renormalization group (NRG)
\cite{wilson1975RenormalizationGroupCritical,peters2006NumericalRenormalizationGroup,bulla2008NumericalRenormalizationGroup}.
NRG can accurately calculate physical quantities such as Green's functions and the self-energy near the Fermi surface at low temperatures.

\section{Results}
\label{section:Results}

\subsection{Effective Hamiltonian and its time-reversal symmetry}
The many-body Hamiltonian of this model fulfills the conventional time-reversal symmetry, defined as
\begin{align}
    T H(\vb*k) T^{-1} = H(-\vb*k),
    \label{eq:conventionalTRS}
\end{align}
where $T = e^{i\pi S_y} \mathcal K$.
$S_y$ is the second quantized spin operator along the $y$ axis, and $\mathcal K$ is the complex conjugation operator.
\begin{figure}[H]
    \centering
    \includegraphics[width = \linewidth]{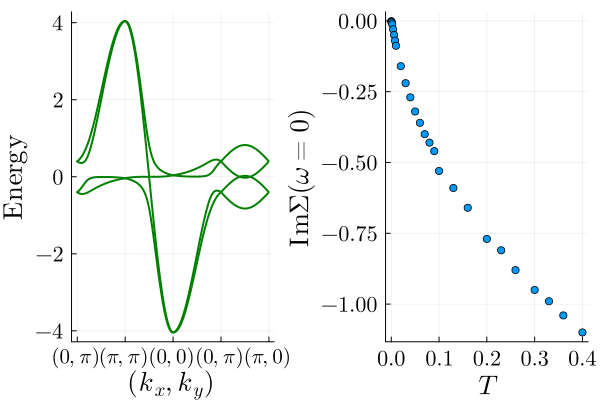}
    \caption{Band structure of the non-interacting model (left) and temperature dependence of the imaginary part of the self-energy for $U=2$ at $\omega =0$ (right). We set $t_c=1, t_f = 0, \mu_c = 0, \mu_f = -U/2, V=0.4, \alpha = 0.3$.}
    \label{fig:2-band_selfenergy}
\end{figure}

 The non-interacting band structure of this model is shown in the left panel of Fig.~\ref{fig:2-band_selfenergy}. The non-interacting band structure consists of a $c$-electron band and an $f$-electron band that are both spin-split due to the SOC. We note that at the time-reversal invariant momenta, the SOC term vanishes in the Hamiltonian, and the spectrum degenerates. These degeneracies play an essential role in the appearance of the NHSE.

Furthermore, the imaginary part of the self-energy at the Fermi energy as calculated by DMFT for different temperatures is shown in the right panel of Fig.~\ref{fig:2-band_selfenergy}.
The magnitude of $\Im \Sigma$ increases with increasing temperature. 
Therefore, we can use self-energy dependence as an equivalent to temperature dependence.
Because only the $f$-electron band includes an on-site interaction, the imaginary part of the self-energy appears only in the diagonal components of the $f$-electron band.

\begin{figure*}[htb]
    \centering
    \includegraphics[width=0.7\linewidth]{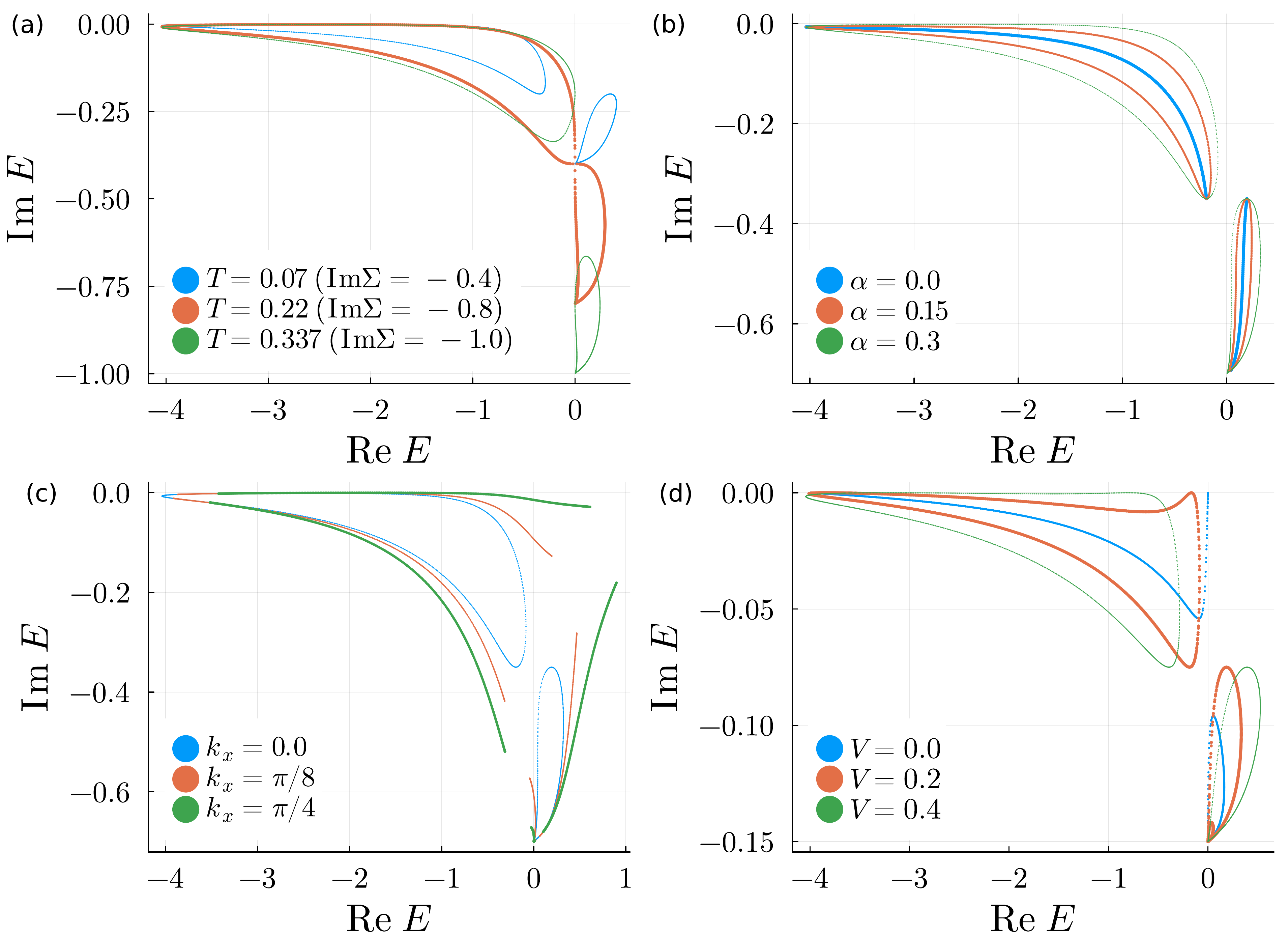}
    \caption{
    (a) The spectrum of $H_{\mathrm{eff}}(k_x, k_y)$ for several values of temperatures at $k_x=0, \pi$.
    (b) The spectrum for several values of SOC, $\alpha$, at $k_x=0, \pi$.
    (c) The spectrum for $k_x=0,\pi/8,\pi/4$.
    (d) The spectrum for several values of $V$. These data are obtained for $t_f = 0$, $t_c = 1.0$, 
    (a) There are two point-gap for high and low temperatures. Around the Kondo temperature, an exceptional point appears, and both bands touch each other.
    (b) Changing the strength of the SOC, we see that the SOC opens the point-gap. The calculation is done for ($\Im \Sigma = -0.7)$ and $V = 0.4$.
    (c) Changing $k_x$, we see that a point-gap only exists for $k_x=0$ and $\pi$. Different values of $k_x$ break the transposed-type time-reversal symmetry resulting in an opening of the previously closed loop.
    (d) Changing the local hybridization $V$, we see that $V$ opens the point-gap of the $c$ electrons. On the other hand, the point-gap of $f$ electrons always opens independently of $V$.}
    \label{fig:2-SpectrumVs}
\end{figure*}

Therefore, we can express the effective Hamiltonian and its eigenvalues as
\begin{align}
    H_{\text{eff}}(\vb*k) &= \mqty(
    \epsilon_{f\vb*k} + i \Im \Sigma & V & 0 & \alpha_{-}(\vb*k) \\
    V & \epsilon_{c\vb*k} & \alpha_{-}(\vb*k) & 0 \\  
    0 & \alpha_{+}(\vb*k) & \epsilon_{f\vb*k} + i \Im \Sigma & V \\
    \alpha_{+}(\vb*k) & 0 & V & \epsilon_{c\vb*k}
    ), \label{eq:EffectiveHamiltonian} \\
    E(\vb*k) &= \frac{\epsilon_{f\vb*k} + i \Im \Sigma + \epsilon_{c\vb*k}}{2} \nonumber \\
    &\pm \sqrt{\pqty{ \frac{\epsilon_{f\vb*k} + i \Im \Sigma - \epsilon_{c\vb*k}}{2}}^2 + \pqty{V \pm \abs{\alpha_{+\vb*k}}}^2},
\end{align}
where 
\begin{align}
    \alpha_{\pm}(\vb*k) = \alpha \sin k_x \pm i \alpha \sin k_y.
\end{align}
From these data, we evaluate the Kondo temperature of this model as the point where the EP appears around the Fermi surface \cite{michishita2020RelationshipExceptionalPoints},
which is equivalent to
\begin{align}
    \epsilon_{c\vb*k} - \epsilon_{f\vb*k} - \Re \Sigma(\omega=0) &= 0, \\
    \Im \Sigma(\omega=0)/2 &= V + \abs{\alpha_{\vb*k}}.
\end{align}
Thus, the Kondo temperature is approximately $T \simeq 0.22t_c$ from Fig.~~\ref{fig:2-band_selfenergy}.

We can define the effective Hamiltonian of the single-particle Green's function
\begin{align}
G(\omega) = (\omega -H_{\mathrm{eff}})^{-1} \qc H_{\mathrm{eff}}=H_0+\Sigma(\omega=0).    
\end{align}
In this paper, we particularly choose $\omega=0$ to discuss the properties of the Fermi energy. 

The time-reversal symmetry of the many-body Hamiltonian [see Eq.~(\ref{eq:conventionalTRS})] imposes the following condition on the single-particle Green's function ~\cite{yoshida2020ExceptionalBandTouching},
\begin{align}
    U_{T} G^\mathrm{T}(\omega+i\delta, \vb*k) U_{T}^{-1} = G(\omega+i\delta, -\vb*k),
\end{align}
using the operator $U_{T} =i \sigma_y \tau_0$. $\sigma_y$ and $\tau_0$ are the Pauli matrices for the spin and orbit, respectively.
The effective Hamiltonian [see Eq.~(\ref{eq:EffectiveHamiltonian})] inherits this symmetry as the transpose-type time-reversal symmetry $(\text{TRS}^\dagger)$, defined as
\begin{equation}
    U_{T} H_{\text{eff}}^\mathrm{T}(k_x,k_y) U_{T}^{-1} = H_{\text{eff}}(-k_x,-k_y).
    \label{eq:TRSdagger}
\end{equation}

In the following, we discuss the parameter dependence of the spectra to clarify the conditions for the non-trivial point-gap topology, which induces the NHSE.
We impose PBC on the $x$ direction and treat $k_x$ as an external parameter.
Therefore, subsystems for each $k_x$ are one-dimensional.
We note that the TRS$^\dagger$ in Eq.~(\ref{eq:TRSdagger}) 
holds only at $k_x=0,\pi$, which are time-reversal invariant momenta, since $k_x$ changes sign under the time-reversal operation.

First, we show the dependence of the spectrum for $k_x=0$ and several values of temperatures in Fig.~\ref{fig:2-SpectrumVs}(a).
At high temperatures, $T \gtrsim 0.22 \  (\Im\Sigma > -0.8)$, the $f$- and $c$-electron bands are split,
forming two separate loops because of the large imaginary part of the self-energy. 
One point-gap is mainly formed by $f$ electrons and has a large imaginary part.
The other point-gap is mainly formed by $c$ electrons and has a small imaginary part.
As the temperature is lowered, the gap separating the $f$- and $c$-electron bands becomes smaller. 
At $T\simeq0.22$ corresponding to $\Im\Sigma=-0.8 = -2V$, an exceptional point appears related to the Kondo effect.
At this temperature, the $c$-electron band and the $f$-electron band touch each other.  
When the imaginary part of the self-energy is small at low temperatures, $T \lesssim 0.22$, both bands form separated point-gap again because of the Hamiltonian's hybridization.

Next, in Fig~\ref{fig:2-SpectrumVs}(b), we show the spectrum of the effective Hamiltonian for $T \simeq 0.174 \ (\Im\Sigma=-0.7)$, $k_x=0$, and different strengths of the SOC.
Without SOC (blue line), the spectrum consists of two lines corresponding to the $c$- and $f$- electron bands, each of which is doubly degenerate since the energy of each $k_y$ is equal to the energy of $-k_y$.
Including the SOC (orange and green lines), we see the opening of a point-gap because the SOC breaks this degeneracy.

Third, we analyze the dependence of the spectrum on $k_x$ in Fig.~\ref{fig:2-SpectrumVs}(c) for $T \simeq 0.174 \ (\Im\Sigma=-0.7)$.
While the spectrum forms closed loops for $k_x=0$, the spectrum forms open lines for $k_x\neq 0, \pi$.
This is caused by the SOC resolving the momentum degeneracy, as explained above.
When $k_x\neq0$, the SOC splits all degeneracies, and the point-gap does not appear (orange and green lines).
However, when $k_x=0$, the SOC does not act on the time-reversal invariant momenta in the Brillouin zone, so the degeneracy partially remains, resulting in a point-gap (blue line).

Finally, the dependence of the spectrum on $V$ is shown in Fig.~\ref{fig:2-SpectrumVs}(d).
There is always a point-gap for the $f$-electron band independent from $V$;
however, the point-gap in the $c$-electron band closes at the limit of $V = 0$.
Therefore, we conclude that the SOC and the imaginary part of the self-energy are sufficient to open a point-gap in the spectrum and that the local hybridization increases the size of the point-gap.

The above analyses indicate that the interplay of the momentum dependence of the SOC and the imaginary part of the self-energy at finite temperatures results in the non-trivial point-gap topology at $k_x=0,\pi$.
In particular, the TRS$^\dagger$ inherited from the interacting many-body Hamiltonian in Eq.~(\ref{eq:TRSdagger}) protects this non-trivial point-gap topology.

\subsection{$\mathbb{Z}_2$-invariant and skin modes}
For $k_x = 0,\pi$, this model belongs to the symplectic class AII${}^\dagger$.
Therefore there exists a $\mathbb{Z}_2$ topological invariant for the point-gap \cite{kawabata2019SymmetryTopologyNonHermitian}, which is defined for each reference energy $E_{\text{ref}}$ as
\begin{align*}
& (-1)^{\nu(E_{\text{ref}})} = \mathrm{sgn} \left[ \frac{\mathrm{Pf} \bqty{(H_{\mathrm{eff}}(k_y=\pi)-E_{\text{ref}})U_T} }{\mathrm{Pf} \bqty{(H_{\mathrm{eff}}(k_y=0)-E_{\text{ref}})U_T} } \right. \\
& \times \left. \exp \bqty{-\frac{1}{2} \int_{k_y=0}^{k_y=\pi} \dd{k_y} \partial_{k_y} \log \det \pqty{H_{\mathrm{eff}}(k_y)-E_{\text{ref}}} U_T} \right].
\end{align*}
The function $\mathrm{sgn}(x)$ is $1$ and $-1$ for $x>0$ and $x<0$, respectively.
$\mathrm{Pf[A]}$ corresponds to the Pfaffian of the skew-symmetric matrix $A$.

\begin{figure}[tb]
    \centering
    \includegraphics[width=\linewidth]{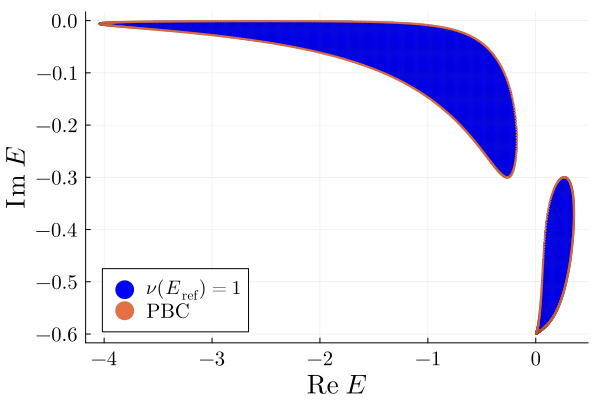}
    \caption{
    $\mathbb{Z}_2$ invariant for different reference  energies at $k_x =0$, $V=0.4, \alpha=0.3$, and  $T = 0.134\,(\Im \Sigma = -0.6)$.
    All wave functions inside the PBC point-gap are topologically non-trivial, as blue corresponds to a negative topological invariant,
    $(-1)^\nu$. The red points correspond to the PBC energies.
    }
    \label{fig:2-Z2inv}
\end{figure}

The result of the topological invariants calculated for $T \simeq 0.174 \ (\Im\Sigma=-0.7)$ is shown in Fig.~\ref{fig:2-Z2inv}. We find that all states inside the point-gap are topologically non-trivial.  The topological invariant is independent of the temperature as long as the imaginary part of the self-energy is finite
[see  Fig.~\ref{fig:2-SpectrumVs}(a)].

\begin{figure*}[bt]
    \centering
    \includegraphics[width=0.9\linewidth]{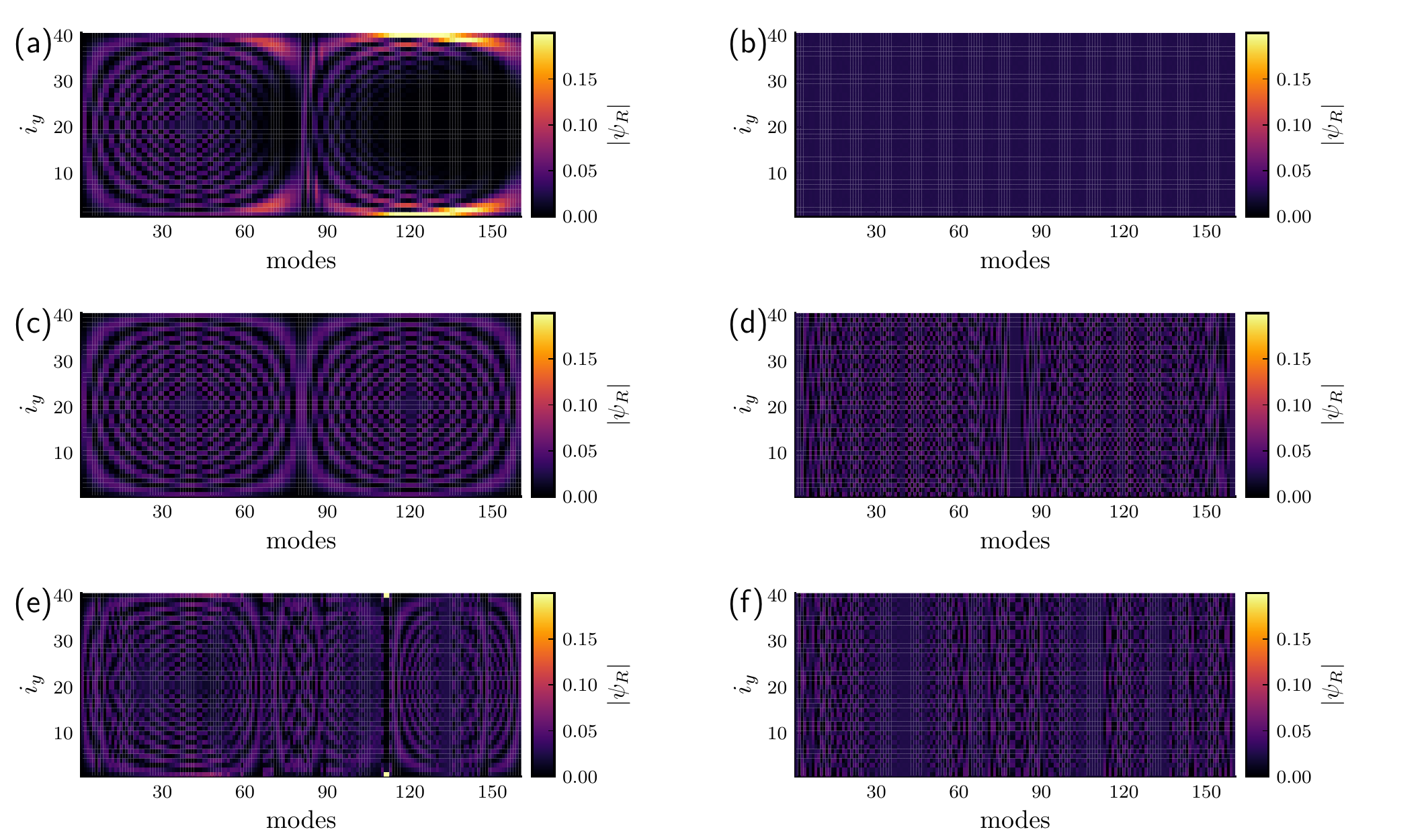}
    \caption{
  The amplitude of the right eigenvectors of the effective Hamiltonian as color over the lattice site in the $y$ direction and the number of the eigenvector:
  The right eigenvectors are calculated for $t_f = 0$, $t_c = 1.0$, $\mu_f = -U/2, \mu_c = 0.0$, $\Im \Sigma=-0.6$, $V = 0.4$, $\alpha=0.3$.
  We use $N=40$ lattice sites in the $y$ direction.
  (a) For $k_x =0$ and OBC in the $y$ direction, the effective one-dimensional subsystem has a transpose-type time-reversal symmetry, and the NHSE occurs.
  While $f$ electrons are strongly localized,  $c$ electrons are weakly localized. 
  (b) For $k_x =0$ and PBC in the $y$ direction, the NHSE does not occur, and all modes are delocalized.
  (c)[(d)] For $\alpha=0, k_x=0$ under OBC [PBC], the subsystem retains the transpose-type time-reversal symmetry but does not have a non-trivial point-gap. Thus, the NHSE does not occur.
  (e)[(f)] For $\alpha=0.3,k_x = \pi/4 (\neq 0)$ and OBC [PBC], the subsystem loses its time-reversal symmetry (i.e., $k_x$ is flipped by the time-reversal operation), and the NHSE does not occur.
    }
    \label{fig:2-wavefunction}
\end{figure*}

The amplitude of the right eigenvectors of $H(k_x)$ calculated for $T=0.274$ $(\Im \Sigma=-0.9)$ depending on the lattice site in the $y$ direction are shown in Fig.~\ref{fig:2-wavefunction}.
Figures~\ref{fig:2-wavefunction}(a) and \ref{fig:2-wavefunction}(b) show the right eigenvectors  for $V=0.4, \alpha=0.3, k_x=0$ under OBC and PBC, respectively.
Under the PBC [see Fig.~\ref{fig:2-wavefunction}(b)], the amplitude of the right eigenvectors is homogeneous.
In contrast, under the OBC [see Fig.~\ref{fig:2-wavefunction}(a)], we see strongly localized modes at the boundaries, as expected when the NHSE occurs.
Especially, the right eigenvectors corresponding to the $f$ electrons shown on the right side are strongly localized, while the $c$ electrons are localized only weakly.
The remaining figures demonstrate the effects of the SOC and the time-reversal symmetry.
Figure~\ref{fig:2-wavefunction}(c) and \ref{fig:2-wavefunction}(e) [\ref{fig:2-wavefunction}(d) and \ref{fig:2-wavefunction}(f)] display the results for $V=0.4, \alpha=0.0, k_x=0$ and $V=0.4, \alpha=0.3, k_x=\pi/4$ under OBC [PBC].
In these figures, most eigenvectors are delocalized for both OBC and PBC, corresponding to the absence of the NHSE.
These facts confirm the results of the energy spectrum shown in the previous section. 

The analyses of the previous section and this section can be summarized as follows.
When the SOC is finite, the effective Hamiltonian has non-trivial point-gap topology induced by the SOC at $k_x = 0,\pi$ at finite temperature
\footnote{
For the case of $k_x\neq 0,\pi$, this subsystem belongs to class AI${}^\dagger$, which destroys the NHSE, with another type of time-reversal symmetry defined as,
$\tilde U_{T} H(k_x, k_y)^T \tilde U_{T}^{-1} = H(k_x, -k_y)$ where $\tilde U_T = 1$.
We emphasize that this symmetry does not originate from the many-body Hamiltonian but arises accidentally.
}, and this point-gap topology induces the $\mathbb Z_2$ NHSE.
We provide a complementary explanation of the $\mathbb Z_2$ NHSE's appearance in Appendix \ref{subsec:appendix_Z2NHSE}.
In the following discussion, we set  $k_x = 0$ for simplicity.

\subsection{Pseudospectrum Analysis}
This section discusses the relationship between the NHSE and physical quantities using the pseudospectrum.
The $\epsilon$-pseudospectrum for a small real number $\epsilon$ is defined as 
\begin{align}
    \sigma_\epsilon (H_{\mathrm{eff}}) &= \bigcup_{\norm{W}_2 < \epsilon} \sigma(H_{\mathrm{eff}} + W), \label{eq:pseudo_spec}\\
    \norm{W}_2 &\equiv \max_{\vb*{v}} \frac{\norm{W\vb*{v}}}{\norm{\vb*{v}}},    
\end{align}
where $\sigma(A)$ denotes the spectrum of operator $A$, and $\norm{\vb*{v}}$ is the conventional 2-norm of the vector $\vb*{v}$;
$\norm{\vb*v}_2 = [\sum_j v_j v_j^*]^{1/2}$.
Thus, the pseudospectrum characterizes the non-normality of the Hamiltonian through the stability of the eigenvalues against small perturbations $W$.
In the thermodynamic limit, the pseudospectrum also satisfies the following property \cite{okuma2021NonHermitianSkinEffects}:
\begin{align}
    \lim_{\epsilon \to 0} \lim_{N \to \infty} \sigma_{\epsilon}\pqty{H_{\mathrm{eff}}^{(N)}} = \overline{\sigma(H_{\mathrm{eff}}(\vb*k))},
    \label{eq:4-PSandPBC}
\end{align}
where $\overline{O}$ denotes the topologically nontrivial part of the closure of $O$.
This means that the NHSE causes a drastic change to the pseudospectrum under infinitesimal perturbation. The pseudospectrum can be an extended set if the spectrum of the Hamiltonian includes a point-gap structure\cite{okuma2021NonHermitianSkinEffects}.
On the other hand, when the NHSE does not occur, including in the case of a Hermitian Hamiltonian, the pseudospectrum is the $\epsilon$ neighborhood of the eigenvalues of the Hamiltonian.
Therefore, the pseudospectrum can detect the appearance of the NHSE.

\begin{figure}
    \centering
    \includegraphics[width=\linewidth]{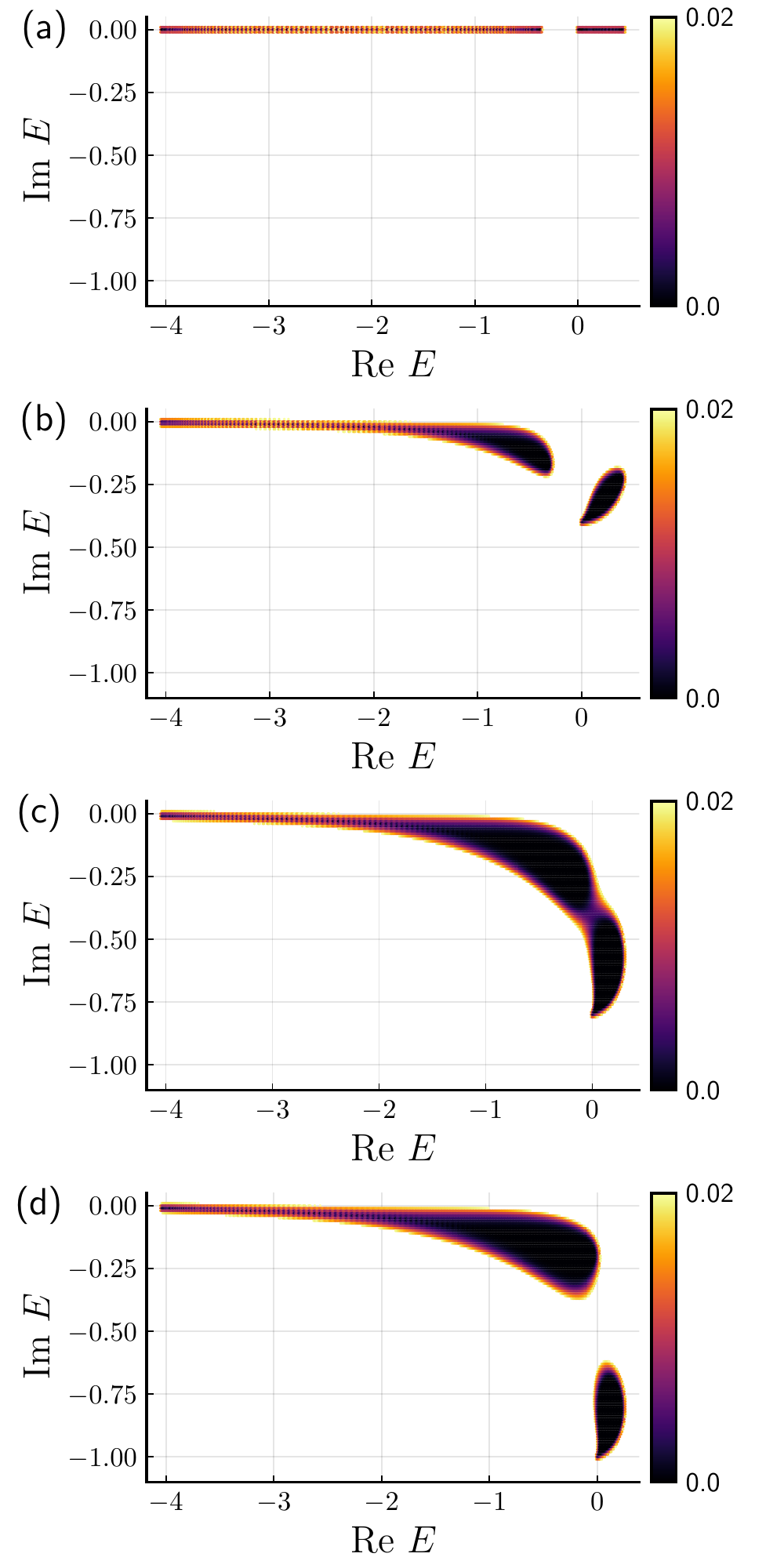}
    \caption{The pseudospectrum at $T = 0$, $0.07$, $0.022$, $0.337$  $(\Im \Sigma = 0$, $-0.4$, $-0.8$, $-1.0)$ calculated with $t_c = 1.0$, $t_f=0.0$, $\mu_c = \mu_f = 0.0$, $V=0.4$, $\alpha=0.3$, $k_x = 0.0$, OBC in the $y$ direction, and $120$ sites: (a) In the Hermitian case, the pseudospectrum corresponds to the $\epsilon$-neighbor of the PBC spectrum. (b)-(d) In non-Hermitian cases, the pseudospectrum is the closure of the PBC spectrum. (b) [(d)] In the low [high] temperature regime, the two bands do not hybridize with each other, and the pseudospectrum also consists of two parts. (c)  Around the Kondo temperature, the appearance of the exceptional point results in a continuous pseudospectrum. The color in the figure indicates the size of the used $\epsilon$.}
    \label{fig:4-PseudoSpectrumvsT}
\end{figure}

We show the pseudospectrum of the effective Hamiltonian at the Fermi energy using OBC in the $y$ direction in Fig.~\ref{fig:4-PseudoSpectrumvsT}.
In the Hermitian cases, the pseudospectrum corresponds to the $\epsilon$ neighbor of the PBC spectrum, shown in Fig.~\ref{fig:4-PseudoSpectrumvsT}(a).
In the non-Hermitian cases, Fig.~\ref{fig:4-PseudoSpectrumvsT}(b) - (d), the pseudospectrum deviates significantly from the PBC spectrum. In these cases, the calculated pseudospectrum corresponds to the entire region inside the point-gap. Furthermore, we see that for the temperature at which the exceptional point appears,
the pseudospectrum of both point-gap are connected [see Fig.~\ref{fig:4-PseudoSpectrumvsT}(c)] and strongly enhanced.
This enhancement should vanish in the limit $\epsilon \to 0$. However, the sensitivity of the EP revives the enhancement even with sufficiently small $\epsilon$.

We characterize this enhancement by using the area of the pseudospectrum in the two-dimensional energy plane, defined as
\begin{align}
    S_{\epsilon} (H_{\mathrm{eff}}) = \int_{\sigma_{\epsilon}(H_{\mathrm{eff}})} 1 \cdot \dd{(\Re E)} \dd{(\Im E)}.
\end{align}

\begin{figure}[bt]
    \includegraphics[width=\linewidth]{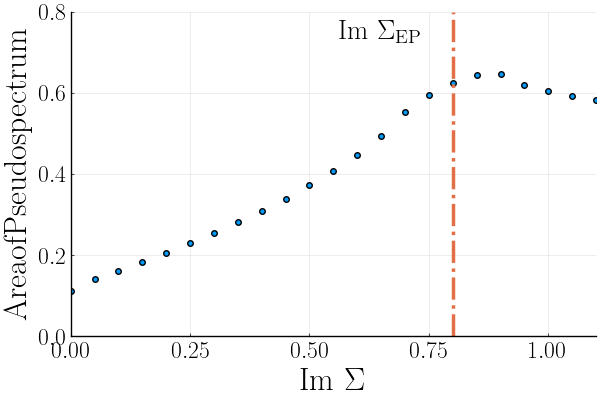}
    \caption{The dependence of the area of the pseudospectrum on the imaginary part of the self-energy: We use $\epsilon=0.02$.
    From Fig. \ref{fig:2-band_selfenergy}, the self-energy dependence can be regarded as a temperature dependence.}
    \label{fig:5-PseudoSpectrum_area}
\end{figure}

The area of the pseudospectrum over the temperature is shown in Fig.~\ref{fig:5-PseudoSpectrum_area}.
At $\Im \Sigma = 0, (T=0)$, the non-trivial point-gap structure disappears, and the pseudospectrum corresponds to the $\epsilon$ neighbor of the PBC spectrum.
Therefore, the area of the pseudospectrum can be approximated as a rectangle of a length four and a height of $2\epsilon$.
Using $\epsilon=0.02$ in Fig.~~\ref{fig:5-PseudoSpectrum_area}, we can estimate $S_{\epsilon}(H) \simeq 0.1$.

The non-trivial point-gap structure appears at finite temperatures and $S_\epsilon$ increases.
It is maximal around the Kondo temperature ($\Im \Sigma = -0.8$), where the EP appears, and both point-gap are connected.
On the high-temperature side, the pseudospectrum is split into two parts again by a large imaginary gap, and there is no EP that enhances the pseudospectrum.
Thus, $S_\epsilon(H)$ becomes smaller.

The above discussion using the pseudospectrum demonstrates the appearance of the NHSE at finite temperatures.
It shows that the nature of the NHSE changes when the EP appears in the energy spectrum.

\subsection{Generalized Brillouin Zone Analysis}
Here, we further study the temperature effects on the NHSE using the generalized Brillouin zone (GBZ).
We determine the localization strength of the wave functions, their temperature dependence, and the impact of the EP.
We use the GBZ theory for symplectic cases here \cite{kawabata2020NonBlochBandTheory}.

In the GBZ theory, the non-Bloch Hamiltonian $H_{\text{nB}}(\beta) = H_{\text{B}}(k_y = -i \log \beta)$,
which is equal to the Bloch Hamiltonian using a complex-valued momentum.
The imaginary part of this complex momentum determines the localization of the wave function.
Thus, we need to find proper $\beta$s that satisfy
\begin{align}
    \det [H_{\text{nB}}(\beta)-E] = 0
\end{align}
for each $E$ in the OBC spectrum.
This would be an 8th-order algebraic equation for $\beta$ in the present model.
However, the time-reversal symmetry separates this equation into two parts,
\begin{align}
    \det [H_{\text{nB}}(\beta) - E] &= g(\beta,E) g(\beta^{-1},E) = 0
    \label{eq:3-detHdecomp}
\end{align}
where
\begin{align}
g(\beta,E) &= \frac{\alpha^2}{4} \beta^2 + \bqty{E + i\pqty{\Im \Sigma  + \alpha V}} \beta \nonumber \\
    &+ \bqty{E^2 + (i \Im \Sigma + 2)E + 2 i \Im \Sigma - \pqty{V^2 + \frac{\alpha^2}{2}}} \nonumber \\
    &+ \bqty{E + i \pqty{\Im \Sigma - \alpha V}} \beta^{-1} + \frac{\alpha^2}{4} \beta^{-2}.
\end{align}
Furthermore, the solution can be improved using the similarity transformation method \cite{wu2022ConnectionsOpenboundarySpectrum}.
The GBZ and the similarity transformation are briefly reviewed in Appendices \ref{subsec:appendix_GBZ} and \ref{subsec:appendix_SimTrans}, respectively.

\begin{figure}[tb]
    \centering
    \includegraphics[width=\linewidth]{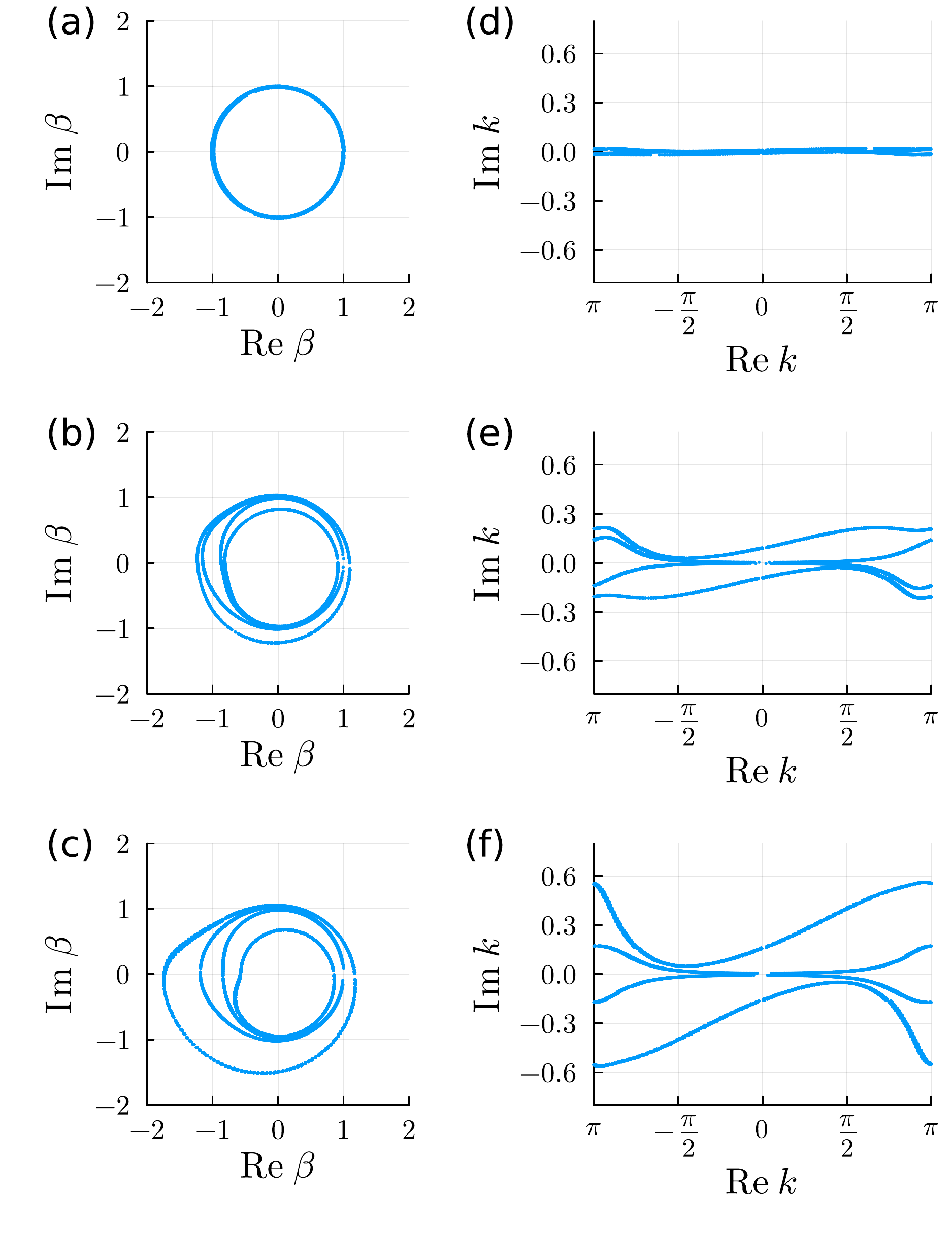}
    \caption{(a)-(c): GBZs on the $\beta$-plane, and (d)-(f) GBZ on the complex $k$-plane $(k=-i \log \beta)$  at $T=0.05,0.099,0.274$,$(\Im \Sigma = 0.05, 0.5, 0.9)$, with $t_c = 1.0, t_f=0.0$, $\mu_c = \mu_f = 0.0$, $V=0.4$,$\alpha=0.3$, $k_x = 0.0$, OBC in the y direction, and $50$ sites. (a)[(d)] For $T \simeq 0$, the Hamiltonian is almost Hermitian, and the GBZ coincides with the 1st BZ. When the NHSE appears, (b),(c),[(e),(f)], the GBZ changes from the unit circle [from the real axis]  with increasing temperature.}
    \label{fig:3-GBZ}
\end{figure}

The results for the current model are shown in Fig.~\ref{fig:3-GBZ}. The left half of the panels show the results on the $\beta$ plane, and the right half of the panels show the results on the complex $k$ plane, where $k=-i \log \beta$. The latter is point symmetric around the origin, as explained above.
When $T=0 \ (\Im \Sigma = 0)$,  the Hamiltonian is Hermitian, and the GBZ corresponds to the unit circle $\abs{\beta}=1$.
Therefore, this model has no localized states for $T=0$.
Increasing the temperature and thus the imaginary part of the self-energy, the effective Hamiltonian becomes non-Hermitian, and the structure of GBZ changes.
As soon as the imaginary part of the self-energy becomes finite, there are solutions for $\beta$ that do not lie on the unit circle [see Fig. \ref{fig:3-GBZ}(a)].
Thus, there will be localized edge modes. We furthermore see that when the imaginary part of the self-energy increases, the deviation from the unit circle increases.
Accordingly, some modes will be strongly localized on the subsystem's edges [see Fig. \ref{fig:3-GBZ}(b)]. 
From Fig.~ \ref{fig:3-GBZ}(d)-\ref{fig:3-GBZ}(f),  we see that the localized modes are mainly around $\Re k=\pi$.
Furthermore, the temperature dependence around $\Re k=\pi$ is also strong.

\begin{figure}[bt]
    \includegraphics[width=\linewidth]{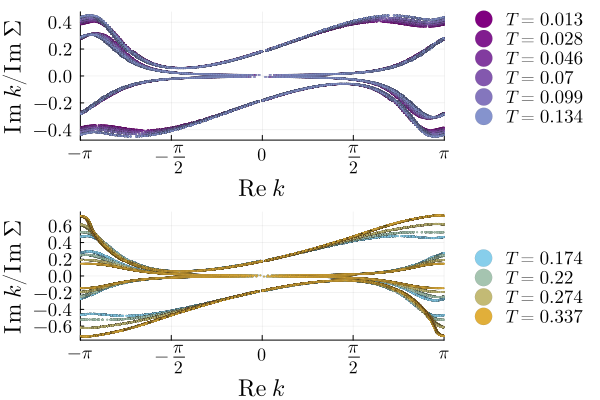}
     \caption{Scaling law of the GBZ on the complex $k$-plane: At low temperatures, $T \lesssim 0.10 \  (\Im \Sigma \lesssim 0.5)$, the scaling law $\Im k / \Im \Sigma$ is well established. However, this law begins to break down close to the Kondo temperature around $k = \pi$.}
    \label{fig:3-CK_scaling}
\end{figure}

Moreover, we find a peculiar difference in the GBZ between low and high temperatures. When we divide the imaginary part of the complex wavenumber by the value of the self-energy, we find that for low temperatures, all curves of the GBZ lie on top of each other, i.e., $\Im k / \Im \Sigma$  is independent of the temperature and $\Im \Sigma$.
This is shown in Fig.~\ref{fig:3-CK_scaling}.
This linear dependence $\Im  k \sim \Im  \Sigma$
is well established at low temperatures, $T \lesssim 0.10$ $ (\Im \Sigma \lesssim 0.5)$.
However, this law breaks down around the Kondo temperature starting at $\Re k = \pm \pi$.

\begin{figure}[tb]
    \centering
    \includegraphics[width=\linewidth]{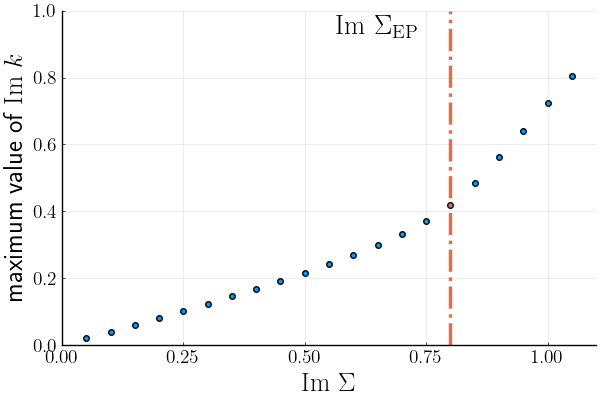}
    \caption{The dependence of the maximum localization of the wave function on the imaginary part of the self-energy: For temperatures below the occurrence of the EP, $(\abs{\Im \Sigma} \lesssim 0.8)$, the maximum $\Im \ k$ increases linearly to the $\Im \ \Sigma$. Above the EP, i.e., above the Kondo temperature, the localization strength increases more quickly than below the Kondo temperature.}
    \label{fig:GBZ-Max_Localization}
\end{figure}

Finally, we plot the wavenumber with the largest imaginary part, i.e., the one with the strongest localization
in Fig.~\ref{fig:GBZ-Max_Localization}. We see that the localization strength  changes around the Kondo temperature.
In particular, most modes are only weakly localized at low temperatures.
Below the Kondo temperature, the localization strength increases linearly.
On the other hand, above the Kondo temperature, the behavior of the localization strength changes and increases much stronger.
This is shown in Fig.~\ref{fig:3-CK_scaling}, where the linear dependence of $\Im k \sim \Im \Sigma$ breaks down around the Kondo temperature.

\section{conclusions}
\label{section:conclusion}
We have demonstrated in this paper that the two-dimensional periodic Anderson model with time-reversal symmetry exhibits the $\mathbb{Z}_2$ NHSE of quasi-particles.
The interplay between the SOC and the finite lifetime of quasi-particles induces the non-trivial point-gap topology, which results in the $\mathbb{Z}_2$ NHSE. 
We have also analyzed the temperature effects on the $\mathbb{Z}_2$ NHSE. 
The area of the pseudospectrum takes a maximum value around the Kondo temperature due to the appearance of an EP in the energy spectrum. 
As the pseudospectrum is related to the sensitivity of the eigenvalues of the Hamiltonian to small perturbations,
this analysis suggests that the single-particle properties are sensitive to external perturbations around the Kondo temperature.

Furthermore, we have analyzed the localization strength of the eigenfunctions by utilizing the generalized Brillouin zone.
We have found that while the localization strength scales linearly with the imaginary part of the self-energy below the Kondo temperature, this scaling law breaks down around the Kondo temperature. 
The localization strength increases much stronger when increasing the imaginary part of the self-energy above the Kondo temperature. 
These changes around the Kondo temperature can be explained by the appearance of an exceptional point in the energy spectrum at the Kondo temperature.

\section*{acknowledgements}
S.K. deeply appreciates the fruitful discussion of Daichi Nakamura, Koki Shinada, Akira Kofuji, and Takahiro Kashihara.
This work is supported by JSPS, KAKENHI Grants No.~JP18K03511 (RP), No.~JP23K03300 (RP), No.~JP21K13850 (TY), and No.~JP22H05247 (TY).

\appendix

\section*{appendix}
\renewcommand{\thesubsection}{\Alph{subsection}}

\subsection{Complementary understanding of the $\mathbb Z_2$ NHSE}
\label{subsec:appendix_Z2NHSE}

We append a discussion on further understanding the occurrence of the $\mathbb Z_2$ NHSE in this model.
In this section, we set $k_x = 0$ for simplicity.
We can  block diagonalize Eq.~(\ref{eq:EffectiveHamiltonian}) as
\begin{align}
    H_{\text{eff}}(\vb*k) &= \mqty(
    H_+ (\vb*k) & 0 \\
    0 & H_- (\vb*k)
    ), \\
    H_\pm (\vb*k) &= \mqty(
    \epsilon_{f\vb*k} + \Im \Sigma & V \pm \alpha \sin k_y \\
    V \pm \alpha \sin k_y & \epsilon_{c\vb*k}
).
\end{align}
We note that the SOC breaks the inversion symmetry, i.e., $U_I H_{\pm}(\vb*k) U_I^\dagger \neq H_{\pm}(-\vb*k)$ ($U_I$ is a unitary operator) for  both block Hamiltonians.
However, the time-reversal symmetry of the effective Hamiltonian, Eq.~(\ref{eq:TRSdagger}), constrains these block Hamiltonians as
\begin{align}
    H_+ (\vb*k) = H_- (-\vb*k). \label{eq:RelOfSubHamiltonian}
\end{align}
When these Hamiltonians have non-trivial point-gap topology, we can define the winding number for each block Hamiltonian as
\begin{align}
W_{\pm}(E_{\mathrm{ref}})=\frac{1}{2\pi i}\int_0^{2\pi}\dd{k_y}\partial_{k_y} \log \det ⁡[H_{\pm}(k_y)-E_\mathrm{ref}].
\end{align}
Equation ~(\ref{eq:RelOfSubHamiltonian}) implies that the winding numbers for each spin sector have opposite signs.

Thus, our model is topologically equivalent to a system, which consists of two subsystems with opposite winding numbers.
These two winding numbers lead to the total system's winding number $W_+(E_{\text{ref}}) + W_- (E_{\text{ref}})$ being 0,
 and the $\mathbb Z_2$ NHSE appears analog to the $\mathbb Z_2$ helical edge mode in Hermitian topological insulators.

\subsection{GBZ theory for the symplectic case}
\label{subsec:appendix_GBZ}
We want to construct the eigenstates of the following Hamiltonian,
\begin{align}
    H = \sum_{n = 1}^N \sum_{j = -J}^J \sum_{\mu,\nu=1}^q (H_j)_{\mu,\nu} c^\dagger_{n + j, \mu} c_{n, \nu},
    \label{eq:3-Hamiltonian}
\end{align}
where $N$ is the system size, $j \in [-J,\cdots,J]$ is the distance of the hopping, and $\mu$, $\nu \in [1,\cdots, q] $ are local degrees of freedom. The Hamiltonian does not need to be Hermitian.

In the conventional Bloch band theory, PBC makes a system discrete-translational invariant.
The Hamiltonian can be block-diagonalized; each block corresponds to a different crystalline momentum.
 Because the crystalline momentum is the eigenvalue of the translational operator, Bloch states are simultaneous eigenstates of the Hamiltonian and the translational operator.
 Furthermore, Bloch's theorem states that wave functions can be decomposed into two parts:
 The part that determines the global behavior of the wave function by the crystal momentum 
 and the part that determines the local behavior having the same periodicity as the crystal.
 The former is written as $e^{ikn}$; the latter is the eigenstate of the Bloch Hamiltonian $H_{\text{B}}(k) = \sum_{j}  H_j e^{-ikj}$.

Using OBC, the system is no longer discrete-translational invariant anymore. Therefore, as defined above, the Bloch states are not generally eigenstates of the Hamiltonian.
However, we can still expand the eigenstates of the OBC system using a linear combination of eigenstates of the translational operator. 
This is called a generalized Bloch state (GBS). The GBS can also be divided into two parts:
a periodic part that is the eigenstate of  $H_{\text{nB}}(\beta) = \sum_{j} H_j \beta^{-j}$,
where $\beta$ is the eigenvalue of the translational operator, and the plane wave part, is written as $\beta^j$.
Under the PBC, the system recovers the translational symmetry, and $\beta^j$ becomes $e^{ikj}$. 
$\beta$ of the GBS is a valuable tool for studying boundary states and analyzing their localization.
For example, the localization of the NHSE wave function is determined by  $\abs{\beta}$.
In particular, if $\abs{\beta}>1$ ($\abs{\beta}<1$), the state is localized at the right (left) edge of the chain. 
Furthermore, the deviation of $\abs{\beta}$ from unity is a measure of the localization strength of the GBS.

In the GBZ theory, the equation determining $\beta$ is
\begin{equation}
    \det [H_{\text{nB}}(\beta) - E] = \beta^{-Jq} f(\beta, E) = 0
    \label{eq:non-Bloch2}.
\end{equation}
where $f(\beta, E)$ is a $ 2Jq$-order algebraic equation determining the $\beta$s.
Because of the boundary condition and the transposed-type time-reversal symmetry, we can find the condition for GBZ as
\begin{align}
    \abs{\beta_{Jq-1}} = \abs{\beta_{Jq}}
    \label{eq:3-conditionforGBZ}
\end{align}
 with $\abs{\beta_1} \le \cdots \le \abs{\beta_{Jq}} \le 1 \le \abs{\beta_{Jq+1}} \le \cdots \le \abs{\beta_{2Jq}}$ \cite{kawabata2020NonBlochBandTheory}.
 
When the system fulfills the transposed-type time-reversal symmetry, then if $\beta$ is a solution,  $\beta^{-1}$ is also a solution;
the GBZ is point symmetric for the origin if displayed using complex wavenumbers, $k = -i \log \beta$.
The imaginary part of a complex wavenumber represents the localization strength.

Therefore, we have to do the following steps:
(1) calculate the OBC spectrum $E$,
(2) solve Eq.~(\ref{eq:non-Bloch2}) for each $E$, and (3) determine the GBZ by Eq.~(\ref{eq:3-conditionforGBZ}).

\subsection{Similarity Transformation}
\label{subsec:appendix_SimTrans}
Unfortunately, the spectrum includes finite-size effects and large numerical errors if we calculate the GBZ using Eq.~(\ref{eq:non-Bloch2}) with OBC. 
We can avoid these problems by using the similarity transformation. Following Ref.~\cite{wu2022ConnectionsOpenboundarySpectrum}, we define this similarity transformation
\begin{align}
    S_{\rho} &= \text{diag}(1, \rho, \rho^2, \cdots, \rho^{N-1}) \\
    [H]_\rho &= S^{-1}_\rho H S_\rho
\end{align}
Under this transformation, the non-Bloch Hamiltonian is changed as
\begin{align}
   \bqty{H_{\text{nB}} (\beta)}_\rho =  \sum_j H_j \rho^{-j} \beta^{-j} = H_{\text{nB}}(\rho \beta).
\end{align}

This transformation results in the most important relation between the OBC spectrum and the PBC spectrum,
$H_{\text{nB}} (\rho e^{ik}) = \bqty{H_{\text{B}} (k)}_\rho$.
This relation can be used to obtain the OBC spectrum from the transformed PBC spectrum.

We can find the selection criterion for $E$
by using the following winding number\cite{wu2022ConnectionsOpenboundarySpectrum} 
\begin{align}
    w_\rho (E) &= \frac{1}{2 \pi i} \int_{0}^{2\pi} \dd k \pdv k \log \det \bqty{ \bqty{H_{\text{B}} (k)}_\rho -E } \nonumber \\
    &= \frac{1}{2 \pi i} \oint_{\abs{\beta}=\rho} \dd \beta \pdv \beta \log \det \bqty{ H_{\text{nB} } (\beta)-E } \nonumber \\
    &= N_\rho (E) - Jq,
\end{align}
where $N_\rho (E)$ corresponds to the number of $\beta (E)$s determined by Eq.~(\ref{eq:non-Bloch2}) surrounded by a circle in the $\beta$ plane, $\abs{\beta} = \rho$.
Using Eq.~(\ref{eq:3-conditionforGBZ}), the spectrum should have crossing points as $E(\beta_{Jq-1})=E(\beta_{Jq})$.
Around a crossing point, the winding number should be $w_\rho (E) = 0/\pm 1/\pm 2$.
Thus, we can determine the GBZ by searching the crossing points in the spectrum of $\bqty{H_{\text{B}} (k)}_\rho$ for each $\rho \in (1,\infty)$, and calculate the corresponding winding numbers.

Finally, we note a few points:
When all skin modes are localized equivalently, the imaginary gauge transformation is a valuable tool for estimating the localization scale \cite{Lee2019AnatomyOfSkinModesAndTopology}.
On the other hand, in a situation where the localization scale varies depending on the mode, we must use the GBZ combined with this similarity transformation method, which is the extension of the "momentum-dependent" imaginary gauge transformation.
The similarity transformation method can calculate OBC spectra and the GBZ with high efficiency and accuracy even when the GBZ has a complicated structure but does not work when the GBZ coincides with the circle, including in the Hermitian case\cite{wu2022ConnectionsOpenboundarySpectrum}.
We must use the "momentum-independent" imaginary-gauge transformation in such cases.

\bibliography{paper}

\end{document}